\documentclass[useAMS,usenatbib]{mn2e}

\usepackage{graphicx}
\usepackage{amsmath}
\usepackage{amsfonts}
\usepackage{amssymb}

\def\ncl{\mbox{$n_{cl}$}}
\def\fcl{\mbox{$f_{cl}$}}
\def\nbg{\mbox{$n_{bg}$}}
\def\fbg{\mbox{$f_{bg}$}}
\def\Fcl{\mbox{$F_{cl}\;$}}
\def\Fbg{\mbox{$F_{bg}\;$}}

\title[The tidal tails of NGC 2298]{The tidal tails of NGC 2298}

\author[E. Balbinot et al.]{Eduardo Balbinot$^{1,2}$\thanks{e-mail: balbinot@if.ufrgs.br}, Bas\'ilio X. Santiago$^{1,2}$, Luiz N. da Costa$^{2,3}$,
\newauthor Martin Makler$^{2,4}$, and Marcio A. G. Maia$^{2,3}$\\
$^1$Departamento de Astronomia, Universidade Federal do Rio Grande do Sul, Av. Bento Gon\c{c}alves 9500, Porto Alegre 91501-970, RS, Brazil \\ 
$^2$Laborat\'orio Interinstitucional de e-Astronomia - LIneA, Rua Gal. Jos\'e Cristino 77, Rio de Janeiro, RJ - 20921-400, Brazil \\ 
$^3$Observat\'orio Nacional, Rua Gal. Jos\'e Cristino 77, Rio de Janeiro, RJ - 22460-040, Brazil \\
$^4$Centro Brasileiro de Pesquisas F\'isicas, Rua Dr. Xavier Sigaud 150, Rio de Janeiro, RJ - 22290-180, Brazil}

\begin{document}

\pagerange{\pageref{firstpage}--\pageref{lastpage}} 

\maketitle

\label{firstpage}

\begin{abstract}
We present an implementation of the matched-filter technique to detect tidal tails of
globular clusters. The method was tested using SDSS data for the globular cluster Palomar 5 
revealing its well known tidal tails. We also ran a simulation of a globular cluster with a tidal tail where
we successfully recover the tails for a cluster at the same position and with the same characteristics of
NGC 2298. Based on the simulation we estimate that the matched-filter increases the contrast of the tail relative to
the background of stars by a factor of $2.5$ for the case of NGC 2298.
We also present the photometry of the globular cluster NGC 2298 using the MOSAIC2 camera installed on the CTIO
4m telescope. The photometry covers $\sim 3 deg^2$ reaching $V\sim23$. A fit
of a King profile to the radial density profile of NGC 2298 shows that this
cluster has a tidal radius of $15.91\arcmin  \pm 1.07\arcmin$ which is twice as in the literature. The
application of the matched-filter to NGC 2298 reveals several extra-tidal structures, including
a leading and trailing tail. We also find that NGC 2298 has extra-tidal structures stretching 
towards and against the Galactic disk, suggesting strong tidal interaction. 
Finally, we assess how the matched-filter performs when applied to a globular cluster 
with and without mass segregation taken into account. We find that disregarding 
the effects of mass segregation may significantly reduce the detection limit of
the matched-filter.
\end{abstract}


\section{Introduction}

Globular clusters (GCs) are the oldest objects found in our Galaxy, hence 
they witnessed the early formation of the Milky Way (MW). Throughout the 
existence of a cluster, it looses stars by a series of both internal and 
external dynamical processes. To understand how stars formed inside a star 
cluster are delivered to the host galaxy is to understand a major part of the 
galaxy formation process in the hierarchical assembly paradigm. 
In that sense, the GCs that we see today are the reminiscent of a much larger 
population of building blocks of our Galaxy.

A GC may loose stars by a number of processes. Their internal dynamics, 
ruled by two-body relaxation, makes stars gradually leave the cluster and leads 
to their eventual dissolution in a time-scale of a few hundred relaxation 
times \citep{binney}. The external influence of the gravitational field of the 
Galaxy may accelerate the dissolution process \citep{spitzer}. The external
field has strong effects over the overall structure of the clusters. One of 
the most clear evidences of this influence is the existence of a limiting 
radius \citep{trager,king68}. \citet{baumgardt} showed that the presence of a 
tidal field throughout the evolution of a cluster results on dramatic
changes on the mass function. This phenomenon was latter observed on several 
clusters \citep{andreuzzi,marchi2298,balbinot09} and is
associated with clusters that are subject to extreme tidal interactions.

While orbiting the host galaxy, a GC experiences a slowly varying external 
potential, which has little effect on its structure, except when crossing 
the disk or bulge of the galaxy. On the crossing event the GC potential is 
rapidly changed, shrinking the tidal radius in a time-scale shorter than the 
cluster dynamical time, rapidly turning bound stars into unbound ones. This
creates a preferential way of scape along the line of action of the tidal 
forces. Stars that leave through the inside of the GC orbit will leap 
forward in the cluster path and stars on the outside will lag behind in 
the orbit. Since the velocity dispersion of the stars in the cluster is much 
less than the orbital velocity of the cluster, the stars that become loose 
follow approximately the same orbit. We may think of each unbound star
as a test particle for the gravitational potential of the MW. Thus, by 
finding which orbit solution best fits the observed tail distribution, we 
may infer the best model for the MW potential \citep{koposov10}.

The study of tidal tails necessarily requires a photometrically homogeneous 
dataset of a large number of stars to the faintest magnitudes possible. There 
were attempts to find tidal structures on several clusters using photographic 
plates \citep{leon}, finding only mild evidences of tidal structures in 20 
GCs. More crucial to tidal tails analyses is the need of a large enough 
solid angle, since the tails may extend over tens of degrees on the sky. 
With the release of the Sloan Digital Sky Survey \citep{york}, it
was possible to investigate large areas of the sky with deep and accurate 
photometry. SDSS led to many discoveries such as tidal streams from disrupting
satellite galaxies \citep{koposov10}, new satellite galaxies \citep{walsh,koposov08}, and tidal tails around GCs 
\citep{rock,oden,grillmair}. 

New large area photometric surveys are being planned for the near future. 
Among them is The Dark Energy Survey (DES). DES is a $5000 deg^2$ photometric 
survey that will cover the southern galactic cap in five filters ($grizY$)
\citep{depoy,mohr}. To achieve this area coverage DES will use a large field of
view camera with an array of 64 high near infra-red efficient CCDs. This new
instrument will be placed at the CTIO Blanco 4 meter telescope. 
DES will reach fainter magnitudes than SDSS with comparable area coverage. 
Although its primary goal is the determination of cosmological model 
parameters, a by-product of DES will be the sampling of stars from our Galaxy, 
which may have great impact over stellar population and Galactic structure 
studies \citep{rossetto}.

In this paper, we develop and validate an implementation of the matched-filter 
technique to detect sparse simple stellar populations, such as GC tidal tails. 
We perform the validation on two controlled scenarios: ($i$) Realistic simulated GC plus a tidal tail; ($ii$) the halo globular cluster Palomar 5, which has a 
previously detected tidal tail. We then apply the algorithm to detect tidal 
structures on the halo globular cluster NGC 2298 which is believed to be a 
cluster on advanced stages of dissolution \citep{marchi2298}. 
Our ultimate goal is to apply the code that we present here to
the entire DES sample and, as a consequence, obtain a homogeneous sample of
such tidal features across the Southern sky. In Sect. 2 we describe the 
matched-filter method. In Sect. 3 we present the validation tests. 
In Sect. 4 we present the NGC 2298 data reduction and analysis of its 
structure and tail. In Sect. 5 we address the impact of mass segregation
over the recovered tail from the matched-filter. In Sect. 6 we present 
our final discussion and conclusions. 

\section{Matched Filter}
\label{findgama}

The matched-filter (MF) is a long used technique developed for signal 
processing \citep{wiener}. The MF technique has a wide field of applications 
in astrophysics going from the detection of clusters of galaxies \citep{kepner} to the characterization of light curves of stars with eclipsing exoplanets \citep{doyle}.

In this work, the MF is used to detect low-density simple stellar 
populations (SSP) that are projected against the Galactic field stars. This 
is done by determining the surface density of
stars that are consistent with a given SSP by means of a weighted least-squares
fit to a carefully constructed model. The implementations of the MF follow 
closely the work of \citet{rock} and \citet{oden}. Although the MF has been
well developed in these previous works, in this work we judge it necessary to 
redescribe the method in face of some additional features that we propose,
which depend on the very definition of the functions and models adopted. 

We want to detect a SSP overlaid with the Galaxy field populations. A simple 
model for the number of stars at a given position ($\alpha, \delta$) as 
a function of colour ($c$) and magnitude ($m$) may be written 
as:

\begin{equation}
N(\alpha,\delta,c,m) = n_{cl} + n_{bg}
\label{eq1}
\end{equation}

where \ncl is the number of stars belonging the SSP and \nbg is the
number of Galaxy field stars. \ncl can be obtained both from observational 
data or from simulations using known properties of this SSP (age, metallicity, 
redenning, distance, mass function, unresolved binary fraction and 
photometric errors).

\ncl$~$may be normalized by the total number of stars that contribute to
the SSP:

\begin{equation}
n_{cl}(\alpha,\delta,c,m) = \zeta_{cl}(\alpha,\delta) f_{cl}(\alpha,\delta,c,m)
\label{ncl}
\end{equation}

where \fcl may be thought of as a probability function as in 
common statistics. In essence, \fcl describes
the probability of randomly drawing a star from the SSP at a given colour 
and magnitude. 

The same procedure may be applied to \nbg.

\begin{equation}
n_{bg}(\alpha,\delta,c,m) = \zeta_{bg}(\alpha,\delta) f_{bg}(\alpha,\delta,c,m)
\label{nbg}
\end{equation}

A simple assumption here is to consider \fcl constant across all analysed field.
This assumption bears some approximations with it. One of them 
is to consider the SSP at the same distance everywhere. Often the tail 
extends through $kpc$ scales, which may lead to a variations on the distance 
modulus with position on the sky. A further 
approximation is to assume that the Present Day Mass Function (PDMF) from 
the cluster is the same as in the tidal tail. \citet{baumgardt} showed that 
dynamically evolved globular clusters have a rapidly evolving stellar mass 
function. So, the stars that are left along the tail may not be well 
described by the cluster PDMF, since these stars left the cluster
on the past in an epoch when the mass function was different. This issue is 
further aggravated by mass segregation, since stars that leave the cluster are 
near its tidal radii, thus having a lower mass than the bulk of stars 
\citep{koch}. In this work, we initially drop the spatial dependency of \fcl, 
leaving an assessment of the impact caused by mass segregation to 
\S \ref{fclmasseg}.

The number of Galaxy field stars is expected to vary slowly on large scales. 
This variation should be reflected on a position dependency on \fbg. The 
scale of the analysed field and its complexity will determine if the spatial 
dependence of \fbg may be disregarded or not. At each case a prescription of 
how the spatial variability was dealt with will be presented.

Considering all approximations quoted above, we are left with a simple model 
for the number of stars as a function of position, colour, and magnitude. 
This model assumes that there are only two stellar populations, the SSP itself
and field stars from the Galaxy. 

\begin{equation}
N(\alpha,\delta,c,m) = \zeta_{cl}(\alpha,\delta)f_{cl}(c,m) + \zeta_{bg}(\alpha,\delta) f_{bg}(\alpha,\delta,c,m)
\label{Npre}
\end{equation}

The construction of \fcl and \fbg is done by means of a Hess diagram, where
we divide the colour-magnitude diagram (CMD) in bins of 0.01 in colour and 0.1 
in magnitude. The resulting diagram is then smoothed using a Gaussian kernel.
We label the CMD bins by the index $j$. The sky is also divided in bins of
right ascension and declination and labelled by the index $i$. For instance, 
$f_{bg}(\alpha_i,\delta_i,c_j,m_j)$ is the measure of \fbg at the $i$-th 
bin of spatial coordinates and the $j$-th CMD bin. From here on we use the
notation $f_{bg}(i,j)$ for simplicity.
Since these functions are now discrete functions of coordinates and 
CMD position, we must work with new discrete functions that are integrals
over the CMD and solid angle bins. The discrete model for the number of 
stars is

\begin{equation}
N(i,j) = \gamma_{cl}(i) F_{cl}(j) + \gamma_{bg}(i)F_{bg}(i,j)
\label{Ndis}
\end{equation}
where 
\begin{eqnarray*}
\gamma_{cl,bg}(i) &=& \int_{\Omega_i} \zeta_{cl,bg} ~d\Omega \nonumber \\
F_{cl}(j) &=& \int_{P_j} f_{cl} ~dm\,dc  \nonumber \\
F_{bg}(i,j) &=& \int_{\Omega_i}  \int_{P_j} f_{bg} ~dm\,dc\,d\Omega \nonumber
\end{eqnarray*}
where $P_j$ is the area of the $j$-th pixel in the CMD and $\Omega_i$ is the 
solid angle covered by the $i$-th spatial bin. 

Using this model for the number of stars in any region of the sky, we may 
now use an observed stellar sample and find the best fit to this model by 
means of a lest-squares fit. Let $n(i,j)$ be the observed distribution of 
stars. At the $i$-th position bin the quantity to be minimized is:

\begin{equation}
S^2(i) = \sum_j \dfrac{[n(i,j) - \gamma_{cl}(i) F_{cl}(j) - \gamma_{bg}(i)F_{bg}(i,j)]^2}{\gamma_{bg}(i)F_{bg}(i,j)} ~.
\label{lsq}
\end{equation}

Minimizing equation (\ref{lsq}) and solving for $\gamma_{cl}$ (i.e. $\frac{d S^2}{d \gamma_{cl}} = 0$) we have:

\begin{equation}
\gamma_{cl}(i) = \dfrac{\sum_j~n(i,j) F_{cl}(j)/F_{bg}(i,j)}{\sum_{j} F_{cl}^2(j)/F_{bg}(i,j)}  - \dfrac{\gamma_{bg}(i)}{\sum_{j} F_{cl}^2(j)/F_{bg}(i,j)} ~.
\label{nc}
\end{equation}

In summary, one must plug in $n(i,j)$ on equation (\ref{nc}) to find the best 
estimate of $\gamma_{cl}$, which is in turn the best estimate of the number of 
stars that are consistent with the SSP. 
Equation (\ref{nc}) differs slightly from that found by \citet{rock}, although 
it matches exactly the one found by \citet{oden}. This discrepancy may be due 
to distinct definitions and model constructions.

Note that in equation (\ref{nc}) the denominator for both terms is constant 
at every position if we neglect the spatial dependency of \Fbg (i.e. no $i$ 
dependency). The number of background stars ($\gamma_{bg}$) can then be 
easily estimated from a polynomial fit to regions where we know for sure 
that there is no contribution from the SSP.

Our implementation of the MF was coded mostly using \texttt{Python}, with 
the aid of the \texttt{SciPy} module for signal processing routines. Some of
 the heavy array manipulation was carried out with Fortran and linked 
to \texttt{Python} using \texttt{F2Py}, which is part of the \texttt{Numpy} 
project. Despite being written in a high level language, our implementation
does not require a great deal of computational power or time since most of 
the math is carried out by external routines which are written in C or 
Fortran (all the matched-filter analysis was carried 
out in a low-end desktop machine). We expect to release a public version 
of the code for the community through the Brazilian DES Science Portal 
in the near future.

\section{Validation tests}

\subsection{Simulations}
\label{sec:simulations}

To properly recreate the conditions where globular clusters are found 
(i.e. projected against a background of Galaxy field stars), we must be 
able to simulate field stars in any given direction of the sky. 
In addition,  we must be able to generate realistic stellar populations 
from any given stellar evolution model, taking into account every observational 
effects. To achieve these goals a variety of softwares had to
be employed.

The simulations of the Galaxy stars was done using the TRIdimensional modeL 
of thE GALaxy (\texttt{Trilegal}\footnote{http://stev.oapd.inaf.it/trilegal}) 
by \citet{girarditri}. The \texttt{Trilegal} code simulates the stellar 
content of the Galaxy in any direction of the sky, including contributions 
from the four basic structural components: thin and thick disk, bulge, and 
halo. We refer to the original paper for further details on the code.

To simulate a globular cluster, we use an adaptation of the code
from \citet{kerber}. The original code simulates the CMD of a 
stellar population for a given stellar evolution model, also 
taking into account the effects of unresolved binaries and observational 
errors. We modified the original code in order 
to include positions, by spreading the stars over the sky according to a 
given mass profile (e.g. \citet{king66}). In addition, a tidal tail 
is added using a $1/r$ density decay profile \citet{decay}. Finally, we allow
a position-dependent PDMF in order to incorporate mass segregation effects.

We simulate a globular cluster located at $\alpha_{J2000} = 6^h48^m59^s$ and  
$\delta_{J2000} = -36^{\circ}00\arcmin02\arcsec$, which corresponds to the 
position of NGC 2298, analyzed in \S \ref{2298}. Its structure is described
by a King profile with a core radius $r_c = 0.91 \arcmin = 2.8 pc$ and 
a concentration parameter $c = log(r_t / r_c) = 0.94$. A Padova evolutionary 
model \citep{leo} was chosen with $\log(age(yr)) = 10.10$ and 
$[Fe/H] = -1.98$. The simulated cluster is placed $10.8 kpc$ away
from the Sun with no reddening, for simplicity. We adopt a Kroupa IMF and
choose not to include mass segregation. The adopted fraction of binaries 
for this simulations is of 50\%. The tidal tails extend $1~kpc$ in each 
direction, with a position angle of $45^\circ$ and an angle with the plane 
of the sky of $20^\circ$. The chosen tidal tails width is $16 \arcmin = 50pc$. 
The simulated globular cluster has $\sim 2\,10^4$ stars with 20\% of 
them belonging to the tidal tail. The simulation was carried out using 
the $g$ and $r$ passbands from DES.

A $20 deg^2$ region of the Galaxy field stars was simulated using 
\texttt{Trilegal}. These stars were uniformly spread in a $24 deg^2$ region
around the simulated star cluster.

Photometric errors were added based on SDSS $r$ magnitudes, which is 
consistent with a photometric detection limit of $r\sim23.0$. This limit is 
also consistent with the observations of NGC 2298 (see \S \ref{sec:2298data} 
for details). 

Since we know \textit{a priori} which stars belong to the simulated cluster, 
it is fairly easy to build \Fcl, the same is true for \Fbg. We applied the MF 
to four tidal tails with different densities, 20\% (3200), 10\% (1600), 5\% (800), and 1\% (160) of the total number of simulated cluster stars. This 
was done by randomly removing stars from the simulated tail. 
Furthermore, we compared the MF results with a simpler
method of quantifying the simulated tidal tail, based on simple star
counts. In this alternative method we compute star counts at each spatial bin,
evaluate the expected average background counts, and subtract this later
from the former.

\begin{figure}
   \begin{center}
   \includegraphics[width=0.5\textwidth]{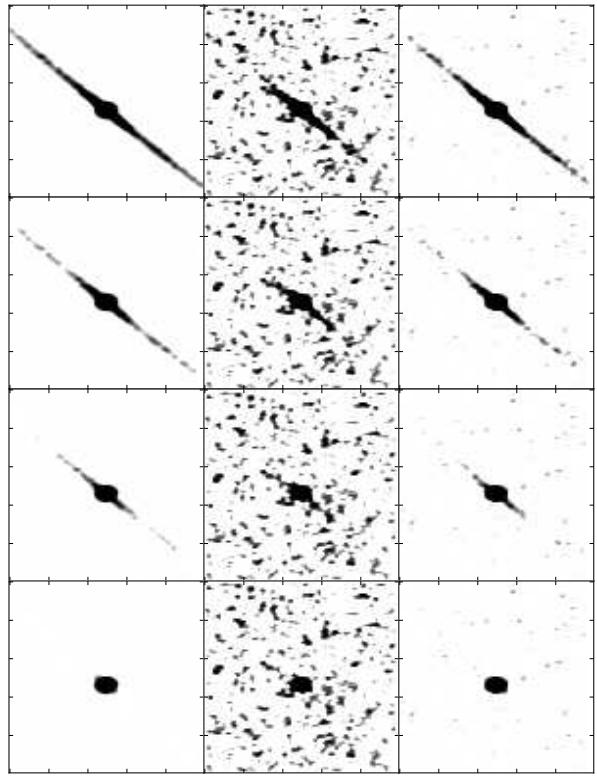}
   \caption{Output of the simulations. Each panel covers $6^\circ$ by $4^\circ$
on the sky. The left column shows the number of simulated cluster and 
tail stars. The central column shows the total number of simulated stars, 
cluster, tail and field, after subtraction of the average number over
the entire simulation field (star counts method). The right column 
shows the output of the matched-filter, $\gamma_{cl}$, as
described in \S \ref{findgama}. }
   \label{sim}
   \end{center}
\end{figure}

\begin{figure}
   \begin{center}
   \includegraphics[width=0.5\textwidth]{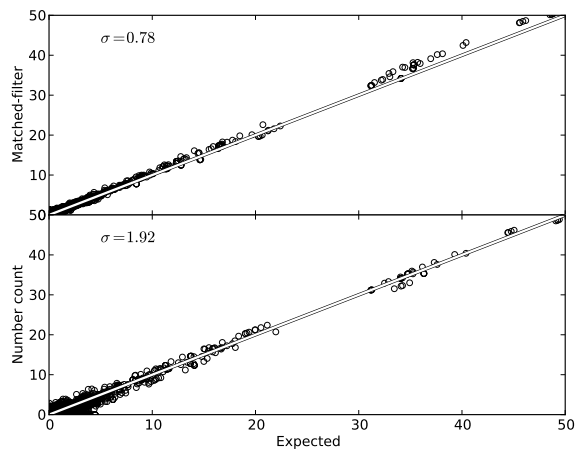}
   \caption{Comparison between the simulated number of cluster stars 
(in the x-axis) with the number derived from simple number counts (lower panel) and from the matched-filter (upper panel). The identity line
   is shown on both panels. The dispersion is also indicated on the top left corner of each panel.}
   \label{matchalpha}
   \end{center}
\end{figure}

Figure \ref{sim} shows the on-sky distribution of simulated stars compared
to the MF results. It is clear that the matched-filter improves the contrast 
of the tail relative to the field stars when compared to a direct star
counting method. Yet, the comparison of the left and right columns
in the figure reveal that the MF does not recover all the structure 
and extension in the tails, specially in the sparser cases. 
In figure \ref{matchalpha} 
we compare, for the two methods, the resulting cluster counts 
to the actual number of simulated cluster stars.
The matched-filter clearly reduces the noise in
this scatter plot by a factor of $1.92 / 0.72 = 2.67$, 
and therefore the contrast with the background, when compared to
simple star counts method. 
We conclude that the our implementation of the matched-filter works well for 
this simulated set.

\subsection{Palomar 5}

In order to further validate our algorithm we have chosen the halo globular cluster Palomar 5. This cluster
has the most prominent tidal tail known to date, making it a good test case for any detection algorithm.

Our analysis was carried out using the Sloan Digital Sky Survey (SDSS) Data Release 7 (DR7) \citep{dr7}. 
SDSS is a large survey, covering up to 10000 square degrees of the northern and part of the 
southern galactic cap, using 5 filters ($ugriz$). Its large continuous area coverage  
and photometric homogeneity make it a very useful data set for the discovery of tidal tails or 
any other large scale sub-structure in the Galaxy.

Although SDSS reaches $r$ magnitudes up to 23.5, we avoid stars fainter than 22.5. This conservative 
limit is set to avoid any complications due to miss-classification of stars and 
galaxies. 

\begin{figure}
   \begin{center}
   \includegraphics[width=0.5\textwidth]{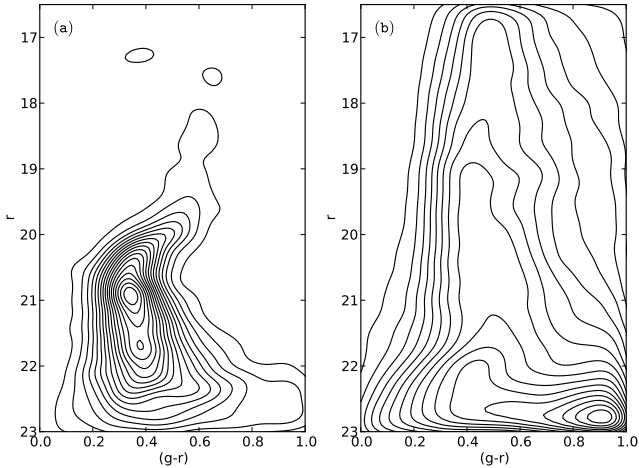}
   \caption{Panel $a$: $r,(g-r)$ \Fcl contours on the CMD plane for the Palomar
5 SDSS data. Panel $b$: same as in panel $a$ but now
   showing the \Fbg contours. In both panels the contour levels are normalized to the range varying from $0$ to $1$, and are evenly spaced spaced by
   $1/20$.}
   \label{fcl_pal}
   \end{center}
\end{figure}

\Fcl was built using a circular region around the center of Palomar 5 with a $0^\circ.13$ radius. 
Figure \ref{fcl_pal}a shows the $r \times (g-r)$ Hess diagram constructed using the stars located inside 
this circular region. Some of the expected features of a typical GC are visible: a main-sequence 
(MS), MS turn-off (MSTO), Red Giant Branch (RGB), and Horizontal Branch (HB). The choice of contour levels
of Figure \ref{fcl_pal} is such that the Asymptotic Giant Branch (AGB), and the Blue Stragglers (BS) are not visible
despite being present in this cluster. To avoid minor contributions of background stars in the region
where \Fcl was built, we only used stars that occupy the loci expected for a GC population.

\begin{figure}
   \includegraphics[width=0.5\textwidth]{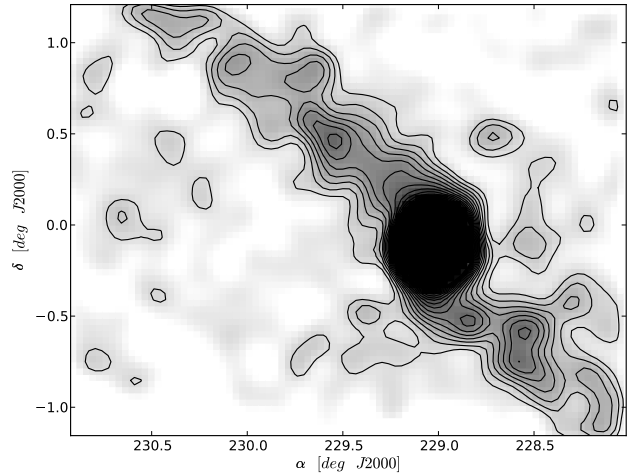}
   \caption{The stellar number density map resulting from the MF 
applied to SDSS data around Palomar 5. We show the number of stars in grey 
scale. The contour map emphasizes the more populated regions. The 4 outermost
contours correspond to 0.1,0.15,0.20,0.25 star $ arcmin^{-2}$. Near the centre of Palomar 5 we do not show any contour for clarity.}
   \label{pal_den}
\end{figure}

As discussed on previous sections, \Fbg is expected to vary over large scales. To accommodate some of this 
variation, we follow the prescriptions of \citet{rock} and take the average Hess diagram of 
four $3~ deg^2$ fields far away from Palomar 5. The fields used are centred in same coordinates as
in \citet{rock}. The resulting Hess diagram is shown in Figure \ref{fcl_pal}b. 
This approach to the construction of \Fbg is such that the spatial dependency should be 
reduced, hence simplifying the solution of equation (\ref{nc}). We thus apply the MF under the assumption that the background 
term does not vary with position. The measured background is of $0.51 \; arcmin^{-2}$. Figure \ref{pal_den} 
shows the smoothed ($0.1 \;deg$ Gaussian smoothing) distribution of stars consistent with Palomar 5 
stellar population.

Having \Fcl and \Fbg properly constructed, we may retrieve the best estimate of the density of stars 
consistent with \Fcl in any region of the sky where \Fbg well describes the 
Galaxy field star population. We applied the matched-filter to a region  
$226^\circ < RA < 231^\circ$ and $-1.1^\circ < Dec < 1.1^\circ$,  which was divided in a grid 
of $0.03 \times 0.03 \; deg$ bins. Figure \ref{pal_den} shows the results of the matched-filter as a stellar surface density 
of Palomar 5 like stars overlaid on a residual contribution by background stars (i.e. the last term on equation \ref{nc}). 

The extra-tidal structure recovered for Palomar 5 closely resembles the 
one found in previous works using the MF technique. The peak of density 
along the tail is expected to be of $\sim 0.2 \; arcmin^{-2}$ according to
\citet{oden}. In this work we find a maximum density of $\sim 0.27 \; arcmin^{-2}$ using  a different set of colours and magnitudes.

The recovered Palomar 5 tail extends throughout $1.8 ~ deg$, ending at 
the edge of the analysed field, which suggests a tail that extends much 
further, as depicted by \citet{oden}. Several density fluctuations are found along the tail. Most of
these were also found on these previous studies. The fluctuations are expected even for the
simplest of the orbits such as circular orbits on a axisymmetric potential \citep{kupper10,kupper08}.

\section{NGC 2298}
\label{2298}

NGC 2298 (also designated by ESO 366-SC 022) is 
located at $l = 245^\circ .63$, $b = -16^\circ .01$, therefore projected towards the
Galactic anti-centre. Its position places it near the Galactic disk. It is thus 
superimposed on to the thin and thick disks, besides the halo. Structural parameters
were found by \citet{marchi2298}, such as core radius $r_c = 0\arcmin .29$ and
tidal radius $r_t = 8\arcmin .0$ leading to a concentration parameter of 
$c = log(r_t/r_c) = 1.44$. The distance and metallicity taken from \citet{harris} are
$d = 10.80\, kpc$ and $[Fe/H] = -1.85$. The analysis of the HST/ACS CMD 
from those authors also yielded an extinction of $E(B-V)=0.15$ towards
the cluster. 

We here describe a first attempt to detect an extra-tidal structure
around NGC 2298, using a field of $\simeq 4$ sq. deg around the cluster.
Its location towards a dense stellar field, with likely varying
extinction, makes it a harsher test to the MF method that we implemented.
NGC 2298 is one of the GCs located in the footprint of DES. 
Therefore, DES will provide a much larger area coverage around the cluster, making it 
possible to make follow-up studies using the same methods developed for this
paper.

\subsection{Data}
\label{sec:2298data}

NGC 2298 was observed using the MOSAIC2 camera located at the 4 meter Blanco 
Telescope at Cerro Tololo International Observatory (CTIO). The MOSAIC2 instrument is a
$8192 px \times 8192 px$ segmented CCD camera with a Field of View (FOV) of $36'\times36'$.  
Each of the 8 camera segments is a CCD with $4096px\times2048px$. Separating each
CCD there is a gap of $35~px = 9.2 \arcsec$ in the East-West direction 
and $50~px = 13.2 \arcsec$ in the North-South direction. The wide field of 
MOSAIC2 makes it 
the best instrument for large area observations in the southern hemisphere. 

\begin{figure}
   \includegraphics[width=0.5\textwidth]{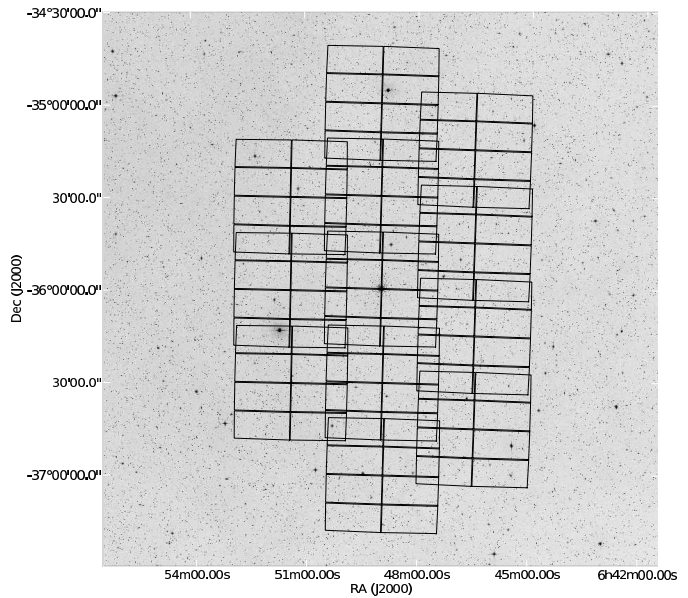}
   \caption{$3\times3 \; deg$ DSS  $r$ band image with the borders of the 12 observed 
   MOSAIC2 FOV overlaid in red. Notice the complex shapes introduced by the gaps between
   the CCDs.}
   \label{observations}
\end{figure}

The observations took place in the night of February 10th 2010 under photometric 
conditions. The mean seeing for the night was $0.7"$, which is normal for the epoch
. We observed 12 overlapping fields around NGC 2298 in two passbands,
V and I. Figure \ref{observations} show the fields and their CCD segments overlaid on a Digital Sky Survey (DSS) image around the cluster. The fields in the Figure
have been corrected for geometric distortions, as explained latter.
The total exposure time was of $240~s$ ($2\times120~s$) in the V band 
and $360~s$ ($3x120~s$) in the I band. The standard stars used for photometric
calibration are taken from 
\citet{stetstd}. They are located within $30'$ of the cluster centre and were 
observed several times during the night, each time using one 
single short exposure in the 
V and I bands. MOSAIC2 was set to $1\times1$ binning on the 8-channel mode. 

The reduction of the data was carried out using the \texttt{MSCRED} package running on 
the \texttt{IRAF} environment. All frames were reduced using standard procedures 
(crosstalk, overscan, bias, flatfield). Some complications arise when dealing
with large field instruments. The FOV of MOSAIC2 introduces a spatial variation 
on the pixel size going from $0.27"/px$ in the centre of the FOV to $0.29"/px$ on 
the edges. This means that each exposure from a given field must be corrected for distortion 
before being stacked, since the projection depends on the pointing of the telescope. 
To correct for distortion each frame must have a reasonably accurate astrometric solution. This
was done by constructing an initial guess for the World Coordinate System (WCS).
This initial guess was determined using USNO-A catalogues
\footnote{Made available to the community at http://www.ctio.noao.edu/mosaic/}. 
With the initial guess for the WCS, we refine the astrometric solution frame by frame using 
the task \texttt{MSCCMATCH}. Frames that will be later combined are registered using
\texttt{MSCIMATCH} and finally, using the task \texttt{MSCIMAGE}, all frames were corrected for 
distortions and combined into a single image. Since the process of distortion correction involves
a re-sampling of the pixels, the bad pixel areas suffer distortions in the process due to the
artificial step discontinuity in the image. To overcome this problem, the bad pixel 
masks were also corrected for distortions and later applied to the final combined image.

\subsection{Photometry}

With the final combined images we performed point spread function (PSF) fit
photometry using the broadly used \texttt{DAOPHOT} software \citet{stetson}. All the 
photometry was performed by an automated python script. The script deals with each 
of the 8 MOSAIC2 chips independently since there may be PSF variations from one chip
to another. The list below shows the steps taken to accomplish the photometry for each
chip.

\begin{enumerate}
\item Find sources above $4~\sigma_{sky}$ (\texttt{DAOFIND}).
\item Run aperture photometry (\texttt{PHOT}).
\item Construct the PSF model using bright non-saturated and isolated stars.
\item Fit the PSF model for each source (\texttt{ALLSTAR}).
\item Transform from physical to world coordinates.
\end{enumerate}

In addition, the PSF was allowed to vary over each chip to account for any residual 
distortions. After this process, we combined the photometric tables from the 
two filters using a positional matching in world coordinates.

The combined VI photometric table for each field was calibrated using the following
calibration equations:

\begin{eqnarray}
V &=& v + a(V-I) + bX + v_0 \nonumber \\
I &=& i + c(V-I) + dX + i_0 \nonumber
\end{eqnarray}

Where $V$ ($I$) is the calibrated magnitude, $v$ ($i$) is the
instrumental magnitude, $(V-I)$ is the calibrated colour, $X$ is the
airmass, and $v_0$ ($i_0$) is the zero-point. The coefficients $(a,b,c,d)$, as
well as the zero-points, were obtained from a fit to the magnitudes and colours of 
the standard stars observed during the night with air-masses ranging from $1.01$ to $2.60$.

The final step to the data reduction is to apply aperture corrections. These corrections
are necessary since there may be {\it seeing} variations from field to field. The aperture corrections
were determined using the overlapping regions on adjacent fields, starting by the central pointing, 
which contains the cluster and the standard stars. 

To eliminate spurious detections, and possibly galaxies, we performed a cut in the magnitude error
of the final photometric table. This cut eliminates sources with errors  
larger than those expected for point sources at their magnitude value. This process eliminates most 
spurious detections including many galaxies that could introduce uncertainties to the
tidal tail detection.

The final photometric sample has approximately $152000$ stars. The mean photometric error
in the range $16 \leq V \leq 22$ is less than $0.05$, which is enough for
our purposes. All magnitudes were corrected for extinction using \citet{schlegel} dust maps. The
mean reddening for the entire observed region is $E(B-V) = 0.20$.

\begin{figure*}[!h]
   \includegraphics[width=0.7\textwidth]{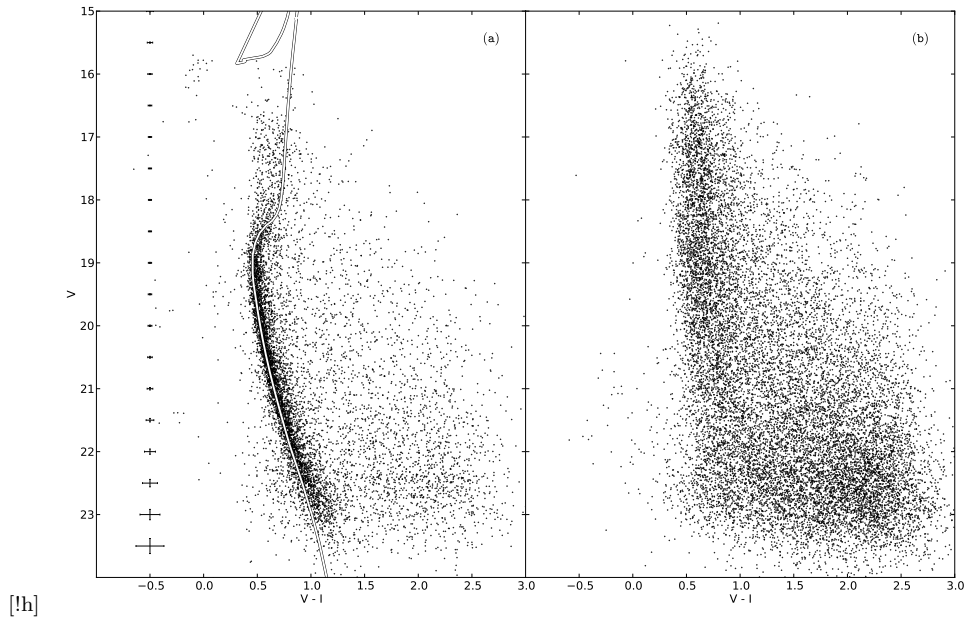}
   \caption{Panel $a$: $V,(V-I)$ CMD of NGC 2298 where only stars inside $r = 37pc$ where
   chosen. The mean photometric error is shown in the extreme left of this panel. We also show the best fit isochrone with $(m-M)=15.15$ and the $(V-I)$ offset of 0.06 explained in the text.
   Panel $b$: $V,(V-I)$ CMD for stars outside $r = 37pc$. Only 10\% of the total stars are shown for
   clarity.}
   \label{cmd2298}
\end{figure*}

In Figure \ref{cmd2298}a we show the CMD for the stars within $12 \arcmin = 
\sim 37~pc$,
which corresponds to 1.5 tidal radius as quoted by \citet{marchi2298}.
The structure of the CMD is typical of an old metal-poor GC. 
Note that at bright magnitudes
we lose stars due to saturation, although the blue end of the extended HB is still visible. Larger
errors on brighter magnitudes are due to saturation in the I band. In Figure \ref{cmd2298}b we
show the CMD for stars outside $12\arcmin$, in this plot we choose to display only a fraction
of $10\%$ of the total number of points for clarity.

Using the best determination of NGC 2298 age and metallicity \citep{marchi2298}, we overlay the corresponding Padova isochrone to the data in Figure \ref{cmd2298}a. The best fit occurs for a distance modulus of $(m-M) = 15.15$,
which closely agrees with the estimate from \citet{harris}.
However, a (V-I) offset of 0.06 towards blue colours was applied to properly fit the isochrone to the MS and MSTO. This offset reflects the
discrepancy between the reddening value from the \citet{schlegel} dust maps,
$E(B-V) = 0.20$, and that found by \citet{marchi2298}.
In what follows, for the sake of coherence, we use the \citet{schlegel}
values over the entire field covered by our MOSAIC2 data.

For completeness reasons we chose not to match stars in overlapping regions. All the analysis
was done on a field by field basis and, when necessary, we adopt the proper area correction (e.g. border
of the fields and chips) using a carefully constructed mask. This mask also takes bad pixel
regions into account when calculating any area. All areas calculated hereafter were 
obtained by a Monte-Carlo integral and accounting for the mask we built. See \citet{balbinot09} 
for further details.

\subsection{Cluster structure}

To make and independent measurement of NGC 2298 structural parameters we built its radial density profile (RDP).
The RDP was built by counting stars in radial bins out to the point where
the background is clearly reached. For the cluster center, we simply
used values given in the literature.

\begin{figure}
   \includegraphics[width=0.5\textwidth]{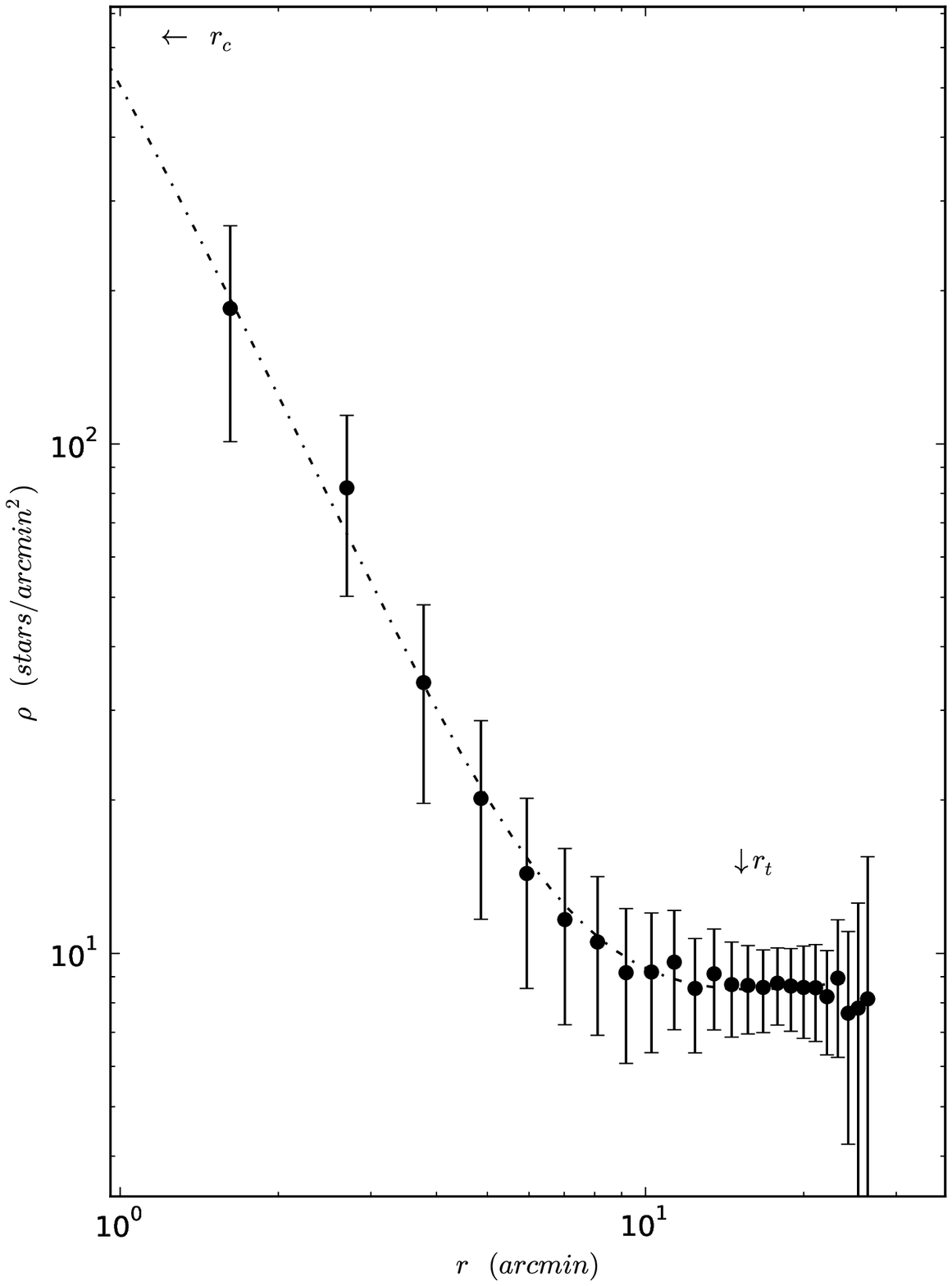}
   \caption{The logarithmic scale RDP for NGC 2298 with 1$\sigma$ error bars.
   The dot-dashed line show the best fit of a King profile with a core
radius fixed at $r_c = 0\arcmin.29 = 0.9 pc$, as found in the high resolution data
from \citet{marchi2298}.
   The best fit values are $r_t=15\arcmin.91\pm 1\arcmin.07$ with a background density of 
   $\sigma_{bg} = 9.50\pm0.10 ~ stars/arcmin^{2}$. The tidal and core radius positions are indicated.
   Notice that the core radius is out of bounds in this plot.}
   \label{rdp}
\end{figure}

In figure \ref{rdp} we show the resulting RDP for NGC 2298. We choose not to use the cluster most central
region due to incompleteness caused both by crowding and gaps on the CCD mosaic. Our RDP analysis covers out to an
angular distance
of $25\arcmin$ which corresponds to $78~pc$. This is approximately three times the best estimated of tidal radius found in the literature. 
We fitted a King profile with fixed core radius ($r_c = 0\arcmin.29 = 0.9pc$) since
our central densities are not accurate due to crowding. 
We find a larger tidal radius of $r_t = 15\arcmin.91 \pm 1\arcmin.07$,
which corresponds to $50~pc$, which is twice larger than previously found. 
The background density found is of $\sigma_{bg} = 9.50\pm0.10 ~ stars/arcmin^{2}$.

The last time NGC 2298 was observed with such a large FOV and photometric depth was by \citet{trager}. The authors find a tidal radius of $6\arcmin.48$ although not reaching deep magnitudes. 
Later, \citet{marchi2298} found a tidal radius of $8\arcmin.0$, although using only 
data that cover a $3.4\times3.4 ~ arcmin$ FOV. Hence their estimated tidal radius
relies on data that do not reach the full extension of the cluster.
Our determination combines the advantages of larger depth compared to
photographic plates and larger FOV compared to HST/ACS.

NGC 2298 was one of the first globular clusters to be found with a high degree
of depletion of low mass stars \citep{marchi2298}. The inverted mass function
and the low concentration parameter are expected for old globular clusters 
that are subject to a high degree of tidal interactions \citet{baumgardt}.

Based on the fitted isochrone, the mass range sampled by our observation
of NGC 2298 is very limited $0.6 \leq M_{\odot} \leq 0.79$. 
The inner parts of NGC 2298 are not accessible to us due to crowding. 
On the outer parts the number of stars is too low at our photometric depth, 
giving low statistical significance to the mass function. We thus refrain from 
making a PDMF reconstruction with our CTIO data; deeper observations covering 
a wider range of masses, such as those that will be provided by DES, are 
necessary to accurately determine the slope of the mass function in
the outskirts of NGC 2298.

\subsection{Extra tidal structure}

We follow the same MF recipe adopted for Palomar 5 in \S \ref{findgama}
in order to investigate the presence of tidal tails associated NGC 2298. 
\Fcl was built using stars that
are less than $37~pc$ from NGC 2298 centre. As seen on figure \ref{cmd2298}, 
a relatively large
amount of field stars are present on the region chosen to build \Fcl. These field 
stars were eliminated by choosing a CMD locus that is consistent with the cluster 
evolutionary sequence in the same fashion as in \citet{balbinot09}. 
\Fbg was built using 6 fields near the edges of the observed region.
\Fcl and \Fbg were built using the completeness limit of the most crowded region. This
conservative approach avoids most issues that might arise from non-homogeneous photometry.

\begin{figure}
   \includegraphics[width=0.5\textwidth]{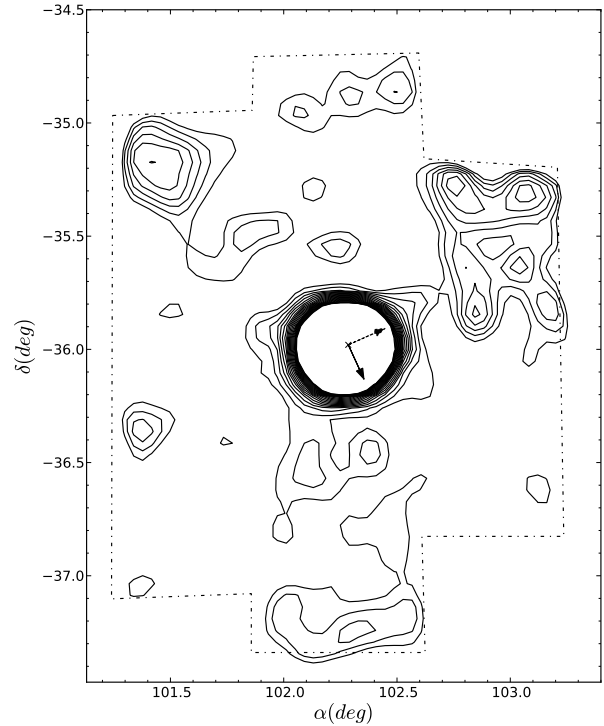}
   \caption{The result of applying the MF to NGC 2298 CTIO data. We show the 
derived number of stars, $\gamma_{cl}$, in a contour plot. The first 
three contour levels correspond to $0.8,1.8,2.8\sigma$ above background. 
   The inner contours start at the tidal radius of the cluster. The dashed arrow points to the 
   direction perpendicular to the Galaxy disk. The solid line points towards the proper 
   motion direction \citep{dinescu}. The resulting map was smoothed using a $0^\circ .06$ Gaussian 
   kernel, thus enhancing structures with a typical size similar to $r_t$.}
   \label{2298tail}
\end{figure}

In figure \ref{2298tail} we show the resulting star count map for NGC 2298 after
applying the MF. Several features are found above the 1$\sigma$ confidence
level. One interpretation of our findings is that the extended Northwest tail
is the trailing tail since its orientation is opposed to the proper motion \citep{dinescu}. The
two smaller opposing structures found in the central East-West direction may be
the result of tidal interaction with the disk, since they point towards
the disk and NGC 2298 is close to the Galactic plane. In addition, a faint structure 
appears ahead of NGC 2298's motion, which may be the leading tail. The two structures 
found at the extreme North and South most likely are boundary effect introduced by the smoothing
process. At last, a strong Northeast structure appears in the direction perpendicular to the
Galactic disk. This may be interpreted as stars that have left the cluster although have not had time
to fall behind or ahead of the orbit. Another faint structure opposite to the previous one is present, although
not connected to any other enhancement. 

Our findings are similar to those predicted by \citet{combes}. The authors simulations predict 
a formation of multiple perpendicular tails in a cross-like pattern resulting from multiple
disk crossings. 

We do not discard the possibility that some of the features detected are in fact due to wrong extinction
corrections. They could result from high frequency structures on the dust filaments close to the disk. These structures 
are not properly sampled due to resolution limitations of the \citet{schlegel} maps built using
IRAS that has a $FWHM = 6\arcmin.1$. However, most of the structures seen in
figure \ref{2298tail} extend along many IRAS FWHMs and thus are likely 
to be real.

\section{Effects of mass segregation}
\label{fclmasseg}

So far, the MF technique has been applied without any assessment of the
influence of improperly built \Fcl. That is, if \Fcl does not reflect
the distribution of cluster stars in the colour-magnitude space, as well
as its variations, throughout the
entire FOV, the application of the MF may lead
to miss identifications of tidal structures. One phenomenon that may
give rise to an improper \Fcl is mass segregation. We here attempt to
quantify its effect on a model cluster consistent with NGC 2298.

To assess what is the effect of mass segregation on the MF technique, 
we made another simulation of a GC similar to NGC 2298, but now 
using the present day mass function
(PDMF) from the literature. Since the slope of the PDMF is only determined
to a distance of $1.8\arcmin$, we extrapolate it to $3.6\arcmin$ by assuming
the same growth rate of the PDMF slope as in the inner parts of NGC 2298.
For the outermost parts of the cluster, we assume that the 
PDMF slope saturates at the value at $3.6\arcmin$. This latter
value, therefore, is the one that applies to most extra-tidal stars
in the model. Table \ref{slopes} lists
the model slopes at different distances from cluster centre.
We also run another simulation, which is identical to the previous one,
but without mass segregation. In this second, non-segregated case, the PDMF
slope used is the one corresponding to the outermost bin in Table \ref{slopes}.

The simulation parameters and number of simulated stars in these
two extra simulations are the same as in 
\S \ref{sec:simulations}, except for the PDMF slopes, as described. The
background of stars is also the same one used in
that section, built using \texttt{TriLegal}.

\begin{table}
\begin{center}
\begin{tabular}{cc}
\hline
Distance ($pc$) & $\alpha$ \\
\hline
\hline
$   0.5      $&$ 1.6   $ \\    
$   1.5      $&$ 1.1   $ \\
$   2.5      $&$ 0.5   $ \\  
$   3.5      $&$ 0.1   $ \\
$   4.5      $&$ 0.0   $ \\
$   5.5      $&$ -0.1   $ \\
$   6.5      $&$ -0.5   $ \\
$   7.5      $&$ -1.1   $ \\
$   8.5      $&$ -1.6   $ \\
$   r\le9.5 $&$  -1.6  $ \\

\hline
\end{tabular}
\caption{PDMF power law slopes for different annuli.
Column 1 shows the distance range in $parsec$. Column 2 shows the simulation
power law slopes for the mass range of $0.08-0.80~M_{\odot}$.}
\label{slopes}
\end{center}
\end{table}

\begin{figure}
   \includegraphics[width=0.5\textwidth]{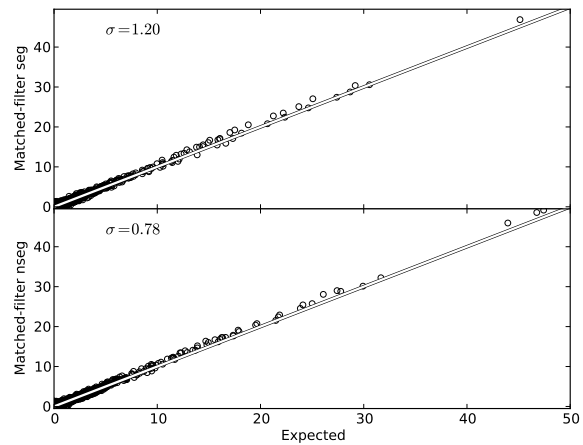}
   \caption{Comparison of the number of cluster stars detected using the
   MF (vertical axis) with the actual number of simulated stars (horizontal axis). The top panel shows the comparison for a mass-segregated cluster, whose
PDMF slopes are listed in Table \ref{slopes}.
   The bottom panel shows the comparison for a non-segregated cluster.
   The identity line is shown on both panels. The dispersion in the
plots are indicated at the top left.}
   \label{finalsim}
\end{figure}

In figure \ref{finalsim} we show the comparison of the true number of cluster
stars to those detected using the MF, in cases when mass-segregation is present
and absent. Notice that
there are no systematic changes in the number of stars obtained in either
situation. However there is
a significant difference on the dispersion, in that the MF reconstructed
star counts have larger scatter in the cluster which is subject to
mass segregation. We thus conclude that, even though the detection of 
a tidal tail is still possible in presence of mass segregation effects, 
the limiting
distance out to which the tail may be detected, as well as some of its
low-density substructure, may be affected if \Fcl does not properly take
segregation into account.

We would like to point out that another
effect that was not taken into account in this work is the difference in 
the completeness of the sample in different regions of the FOV. For instance, 
the cluster core is affected by crowding, thus having a fainter completeness limit than 
the background stars. Using the MF without any completeness correction may
lead to an incorrect surface density of stars, affecting mainly low mass stars. 
One conservative way of resolving this issue is to use the completeness limit of
the most crowded field for all stellar sample analyzed with the MF. Another
way is to run fake stars experiments to properly access the completeness
as a function of colour, magnitude and CCD position.   

\section{Discussion}

We developed an implementation of the MF technique that is relatively
user-independent and with great potential for being used in large scale. 
These features make it 
suitable to be applied to deep and wide angle, such as DES, SDSS, PanSTARSS and
LSST, in a systematic way
to find GC tidal tails and other MW halo sub-structures.

The MF was tested on simulated data, showing that it increases the contrast 
of the tail
relative to the background by a factor of 2.5 for a cluster projected against a dense background, similar to that of a low latitude GC, such as NGC 2298. The MF was also tested on a real scenario were it successfully 
recovered the tidal tail of Palomar 5. 
The Pal 5 tail closely resembles previous detections in the 
literature, reproducing both shape and density.

We then study the GC NGC 2298, which is a good candidate to have a tidal tail due to its
previously studied PDMF and location. We found that the 
cluster has a tidal radius of $r_t=15\arcmin.91\pm 1\arcmin.07$ when a King profile is fitted and keeping the core 
radius fixed ($r_c = 0\arcmin.29$). Our  $r_t$ value is almost twice what was previously found
in the literature. It is based on deeper photometry than previous photographic work and
on a much wider area than previous high-resolution and deep photometry. 
The new value for $r_t$ changes the concentration parameter of NGC 2298 to 
$c = 1.44$, pushing it further away from the $\alpha - c$ relation from 
\citet{marchi}. This discrepancy makes us wonder if the tidal radius of other clusters like NGC 6838 
and NGC 6218 are not similarly affected by observational biases 
associated to small fields or
shallow photometry. {For instance NGC 6218 has a tidal radius of $17.2\arcmin$ whereas observations
only cover $3.4\arcmin$ \citep{marchi06}. There are no publications available for NGC 6838 to properly 
access if its tidal radius determination uses data that extends beyond its literature tidal radius.
We point out that the determination of the tidal radius of Palomar 14 by \citet{Sollima} also shows
an increase by a factor of 4 when compared with to previous determinations in the literature. The
increase of the tidal radius appears to be a trend in the sense that whenever large FOV are used
the tidal radius increases. 

Applying the MF technique to NGC2298, we find that this GC has several extra-tidal structures detected above $1\sigma$ confidence level. The strongest feature
is the elongation of the cluster along the direction of the disk, suggesting strong 
tidal interaction. We also find what appears to be faint leading and 
trailing tails, both extending to the edges of the observed field ($\sim 1^\circ.5$). At
last a large structure is found, spreading from the cluster towards the disk direction. This structure may
be a halo of NGC 2298 that got ejected on the last disk crossing. 

Follow up observations are necessary to properly access the nature of each extra-tidal structure found
around NGC 2298. We expect DES to give this follow up. The survey is going to reach 1 mag deeper 
than our observations with a much larger area, better photometric calibrations, and 3 more passbands. DES will thus allow us to
analyze the extra-tidal structure of NGC 2298 in more detail and over
larger distances.
A larger area will also enable us to test sophistications to the SSP and
background models
underlying the MF, such as a varying background. 

Finally, we simulate a cluster with and without mass segregation to 
evaluate how mass segregation affects the MF results, via an \Fcl that does not 
adequately describe 
the CMD in the outer regions of a mass-segregated cluster.
We find that there is no strong systematic effect in the reconstructed
density of stars. There is, however, a significant difference in the density fluctuations relative to the
truth table, in the sense that the 
the mass-segregated cluster has a larger dispersion around the simulated
densities. This is the result of its
\Fcl not being able to accommodate the variation in the CMD caused by a 
position dependent PDMF. We thus conclude that the impact of unaccounted for mass segregation
in the MF process is to make it more difficult to detect the full extension
and the structural details of a GC tidal tail, rather than preventing
the tidal tail detection per se. From our 
knowledge, this is the first time the MF was tested for an intrinsically
variable and unaccounted for \Fcl.

{\bf Acknowlegments.} 

We are grateful to the CTIO local staff for the help during observation/reduction.
We acknowledge support from Conselho Nacional de Desenvolvimento Cient\'ifico e 
Tecnol\'ogico (CNPq) in Brazil.

We also thank the support of the Laborat\'orio Interinstitucional de e-Astronomia (LIneA) 
operated jointly by the Centro Brasileiro de Pesquisas Fisicas (CBPF), the Laborat\'orio Nacional 
de Computa\c{c}\~ao Cient\'ifica (LNCC) and the Observat\'orio Nacional (ON) and funded by the Ministry 
of Science and Technology (MCT)

{}

\label{lastpage}
\end{document}